\documentclass[letterpaper,english,3p,sort&compress,final]{elsarticle}
\usepackage[T1]{fontenc}
\usepackage[latin9]{inputenc}
\usepackage{prettyref}
\usepackage{amsmath}
\usepackage{graphicx}
\usepackage{amssymb}
\usepackage{esint}

\makeatletter

\providecommand{\tabularnewline}{\\}

\usepackage{graphicx}
\usepackage{dcolumn}
\usepackage{bm}
\usepackage{multirow}
\journal{Nucl. Instrum. Methods Phys. Res., Sect. A}
\def\ps@pprintTitle{%
     \let\@oddhead\@empty
     \let\@evenhead\@empty
     \def\@oddfoot{\footnotesize\itshape
       H.J. Kim, T. Sen, \@journal (2011),
       doi:10.1016/j.nima.2011.03.059\hfill}%
     \let\@evenfoot\@oddfoot}

\@ifundefined{showcaptionsetup}{}{%
 \PassOptionsToPackage{caption=false}{subfig}}
\usepackage{subfig}
\makeatother

\usepackage{babel}

\begin{document}

\title{Beam-beam simulation code \texttt{BBSIM} for particle accelerators}

\author{H.J.~Kim\corref{cor1}}

\ead{hjkim@fnal.gov (H.J. Kim)}

\author{T.~Sen\corref{cor2}}

\cortext[cor1]{Corresponding author. Tel.: +1 630 840 5476; fax: +1 630 840 6039}

\address{Fermi National Accelerator Laboratory, Batavia, Illinois 60510, USA}
\begin{abstract}
A highly efficient, fully parallelized, six-dimensional tracking model
for simulating interactions of colliding hadron beams in high energy
ring colliders and simulating schemes for mitigating their effects
is described. The model uses the weak-strong approximation for calculating
the head-on interactions when the test beam has lower intensity than
the other beam, a look-up table for the efficient calculation of long-range
beam-beam forces, and a self-consistent Poisson solver when both beams
have comparable intensities. A performance test of the model in a
parallel environment is presented. The code is used to calculate beam
emittance and beam loss in the Tevatron at Fermilab and compared with
measurements. We also present results from the studies of two schemes
proposed to compensate the beam-beam interactions: a) the compensation
of long-range interactions in the Relativistic Heavy Ion Collider
(RHIC) at Brookhaven and the Large Hadron Collider (LHC) at CERN with
a current-carrying wire, b) the use of a low-energy electron beam
to compensate the head-on interactions in RHIC.\end{abstract}
\maketitle

\section{Introduction}

In high energy storage-ring colliders, the beam-beam interactions
are known to cause emittance growth and a reduction of beam life time,
and to limit the collider luminosity \citep{Hemery,Evans,Fischer,tsen,zimmermann-2,Shiltsev,Wang}.
It has been a key issue in a high energy collider to simulate the
beam-beam interaction accurately and to mitigate the interaction effects.
A beam-beam simulation code \texttt{BBSIM} has been developed at Fermilab
over the past few years to study the effects of the machine nonlinearities
and the beam-beam interactions \citep{tsen-2,zimmermann,hjkim-2,bbsimc}.
The code is under continuous development with the emphasis being on
including the important details of an accelerator and the ability
to reproduce observations in diagnostic devices. At present, the code
can be used to calculate tune footprints, dynamic apertures, beam
transfer functions, frequency diffusion maps, action diffusion coefficients,
emittance growth, and beam lifetime. Calculation of the last two quantities
over the long time scales of interest is time consuming even with
modern computer technology. In order to run efficiently on a multiprocessor
system, the resulting model was implemented by using parallel libraries
which are MPI (inter-processor Message Passing Interface standard)
\citep{mpi}, state-of-the-art parallel solver libraries (Portable,
Extensible Toolkit for Scientific Calculation, PETSc) \citep{petsc},
and HDF5 (Hierarchical Data Format) \citep{hdf5}.

The organization of the paper is as follows: The physical model used
in the simulation code is described in Section \ref{sec:Physical-model}.
The parallelization algorithm and performance are described in Section
\ref{sec:Parallelization}. Some applications are presented for the
Tevatron, the Relativistic Heavy Ion Collider (RHIC) and the Large
Hadron Collider (LHC) in Section \ref{sec:Application}. Section \ref{sec:Summary}
summarizes our results.

\section{Physical model\label{sec:Physical-model}}

In a collider simulation, the two beams moving in opposite direction
are represented by macroparticles. The macroparticles are generated
with the same charge to mass ratio as the particles in the accelerator.
The number of macroparticles chosen is much less than the bunch intensity
of the beam because it becomes prohibitive to follow approximately
10$^{11}$ particles for even a few revolutions around the accelerator
using modern supercomputers. These macroparticles are generated and
loaded with an initial distribution chosen for the specific simulation
purpose. As an example, a six-dimensional Gaussian distribution is
used for long-term beam evolution. The transverse and longitudinal
motion of particles is calculated by a sequence of linear and nonlinear
transfer maps. During the beam transport, a particle is removed from
the distribution if it reaches a predefined boundary in transverse
or longitudinal direction. In our simulation model, the following
effects are included: head-on and long-range beam-beam interactions,
fields of a current-carrying wire and an electron lens, multipole
errors in quadrupole magnets in interaction regions, sextupoles for
chromaticity correction, ac dipole, resistive wall wake, tune modulation,
noise in lattice elements, single and multiple harmonic rf cavities,
and crab cavities. The finite bunch length effect of the beam-beam
interactions is considered by slicing the beam into several chunks
in the longitudinal direction and then applying a synchro-beam map
\citep{Hirata-1}. Each slice in a beam interacts with slices in the
other beam in turn at a collision point. In the following, linear
and nonlinear tracking models are described in detail.

\subsection{Transport through an arc}

The six-dimensional coordinates of a test particle in the accelerator's
coordinate frame are: $\mathbf{x}=\left(x,x^{'},y,y^{'},z,\delta\right)^{T}$,
where $x$ and $y$ are horizontal and vertical coordinates, $x^{\prime}$
and $y^{\prime}$ the trajectory slopes of the coordinates, $z=-c\Delta t$
the longitudinal distance from the synchronous particle, and $\delta=\Delta p_{z}/p_{0}$
the relative momentum deviation from the synchronous energy \citep{Wille}.
The transverse linear transformation between two elements denoted
by $i$ and $j$ can be written as \begin{equation}
\mathbf{x}_{j}=\left(\begin{array}{cc}
\mathcal{M} & \hat{\mathcal{D}}\\
\hat{\mathcal{A}} & \mathcal{L}\end{array}\right)\mathbf{x}_{i}.\label{eq:t-1}\end{equation}
 Here, $\mathcal{M}$ is a coupled transverse map of \emph{off-momentum}
motion defined by $\mathcal{M}=\mathcal{R}_{j}\tilde{\mathcal{M}}_{i\rightarrow j}\mathcal{R}_{i}^{-1}$,
where $\tilde{\mathcal{M}}_{i\rightarrow j}$ is the uncoupled linear
map described by Twiss functions at $i$ and $j$ elements, and the
transverse coupling matrix $\mathcal{R}$ is defined as \citep{Teng}
\begin{equation}
\mathcal{R}=\frac{1}{\sqrt{1+\left|C\right|}}\left(\begin{array}{cc}
I & C^{\dagger}\\
-C & I\end{array}\right)\label{eq:t-2}\end{equation}
 where $C^{\dagger}$ is the $2\times2$ matrix and the symplectic
conjugate of the coupling matrix $C$. The $4\times2$ dispersion
matrix is defined by $\hat{\mathcal{D}}=\left(0,\mathbf{D}\right)$,
and the dispersion vector $\mathbf{D}=\left(D_{x},D_{x}^{'},D_{y},D_{y}^{'}\right)^{T}$
is characterized by the transverse dispersion functions and the map
$\mathcal{M}$, i.e., $\mathbf{D}=\mathbf{D}_{j}-\mathcal{M}\mathbf{D}_{i}$
where $\mathbf{D}_{i},\mathbf{D}_{j}$ are the dispersion vectors
at $i,j$. Since the transport matrix has to be symplectic, the matrix
$\hat{\mathcal{A}}$ in Eq. (\ref{eq:t-1}) is given by $\hat{A}=-\hat{\mathcal{D}}^{T}S^{T}\mathcal{M},$
where $S$ is a rearranging matrix (see subsection \ref{sub:Symplecticity}).
The longitudinal map $\mathcal{L}$ is given by $\mathcal{L}=\left(\begin{array}{cc}
1 & -\left(\eta/\beta\right)\Delta s\\
0 & 1\end{array}\right)$, where $\eta$ is the slip factor, $\beta=v/c$, and $\Delta s$
the longitudinal distance between the two elements, i.e., $\Delta s=s_{j}-s_{i}$.
It is noted that $s$ is the axis along the beam direction. The nonlinearity
of synchrotron oscillations is applied by adding the longitudinal
momentum change at a rf cavity: \begin{equation}
\Delta\delta=\frac{eV_{rf}}{\beta^{2}E}\left(\sin k_{rf}z-\sin\phi_{s}\right)\label{eq:t-7}\end{equation}
 where $V_{rf}$ is the voltage of rf cavity, $\phi_{s}$ the phase
angle for a synchronous particle with respect to the rf wave, and
$k_{rf}$ the wave number of the rf cavity. If there are higher harmonic
cavities, their effects are added to the momentum change.

\subsection{Beam-beam interactions}

In order to achieve high luminosity in a collider one can increase
the number of bunches which reduces the bunch spacing. More bunches
can increase the number of parasitic encounters in the interaction
regions. Since the calculation of beam-beam forces requires large
amounts of computational resources, it has to be executed rapidly
and accurately. \texttt{BBSIM} has three different models for this
purpose: a weak-strong model for head-on interactions, a look-up table
model for long-range interactions, and a Poisson solver model for
the head-on interactions when both beams have comparable intensities
({}``strong-strong'' model).

\subsubsection{Weak-strong model}

In the weak-strong model we assume that the {}``weak'' beam is affected
by the head-on and long-range interactions while the opposing beam
or {}``strong'' beam is unaffected. The charge distribution of the
strong beam is assumed to be Gaussian: \begin{equation}
\rho\left(x,y,z\right)=\frac{Nq}{\left(2\pi\right)^{3/2}\sigma_{x}\sigma_{y}\sigma_{z}}\exp\left(-\frac{x^{2}}{2\sigma_{x}^{2}}-\frac{y^{2}}{2\sigma_{y}^{2}}-\frac{z^{2}}{2\sigma_{z}^{2}}\right)\label{eq:bb-1}\end{equation}
 Here, $N$ is the number of particles per bunch and $q$ is the charge
per particle. Note that the coordinates $\left(x,y,z\right)$ are
measured in the rest frame of the strong beam. The beam-beam force
between two beams with transverse Gaussian distribution $\rho\left(x,y\right)=\int dz\rho\left(x,y,z\right)$
is well-known \citep{Bassetti}, and the expression for the slope
change is given by, for elliptical beam with $\sigma_{x}>\sigma_{y}$:
\begin{equation}
\left(\begin{array}{c}
\Delta x^{\prime}\\
\Delta y^{\prime}\end{array}\right)=\frac{2Nr_{0}}{\gamma}\frac{\sqrt{\pi}}{\sqrt{2\left(\sigma_{x}^{2}-\sigma_{y}^{2}\right)}}\left(\begin{array}{c}
\text{Im}\left[F\left(x,y\right)\right]\\
\text{Re}\left[F\left(x,y\right)\right]\end{array}\right)\label{eq:bb-2}\end{equation}
 where \begin{equation}
F\left(x,y\right)=w\left(\frac{x+iy}{\sqrt{2\left(\sigma_{x}^{2}-\sigma_{y}^{2}\right)}}\right)-e^{-\frac{x^{2}}{2\sigma_{x}^{2}}-\frac{y^{2}}{2\sigma_{y}^{2}}}w\left(\frac{\frac{x\sigma_{y}}{\sigma_{x}}+i\frac{y\sigma_{x}}{\sigma_{y}}}{\sqrt{2\left(\sigma_{x}^{2}-\sigma_{y}^{2}\right)}}\right).\label{eq:bb-3}\end{equation}
 Here, $w\left(z\right)$ is the complex error function defined by
$w\left(z\right)=e^{-z^{2}}\left(1+\frac{2i}{\sqrt{\pi}}\int_{0}^{z}dt\, e^{t^{2}}\right)$,
and $\gamma$ the Lorentz factor. The constant $r_{0}$ is defined
as $r_{0}\equiv qq_{*}/4\pi\epsilon_{0}m_{0}c^{2}$, where $q_{*}$
is the electric charge of the test particle, and $m_{0}$ the rest
mass of the particle.

\subsubsection{Look-up table model\label{sub:lookup-table}}

The charge distribution of the strong beam in the weak-strong model
is not varied during the simulations. It is redundant to re-calculate
the beam-beam force at every parasitic location and every turn. A
look-up table is one way to avoid it. The look-up table is used to
replace a run time computation with an array indexing operation. The
beam-beam force of a Gaussian beam distribution is described by the
complex error function, as shown in Eq. \prettyref{eq:bb-3}. The
calculation of the complex error function can substantially slow the
beam-beam simulation. However, the look-up table is pre-calculated
and stored in a memory, usually in an array. When the value of the
error function is required, it can be retrieved from the table by
an interpolation scheme, instead of using Eq. \prettyref{eq:bb-3}.
The look-up table method can significantly reduce a computational
cost. The property of the complex error functions yields the symmetry
relations of function $F\left(z\right)$ as \begin{equation}
F\left(-z\right)=-F\left(z\right),\; F\left(\bar{z}\right)=-\overline{F\left(z\right)},\; F\left(-\bar{z}\right)=\overline{F\left(z\right)}\label{eq:lt-1}\end{equation}
 where $z=x+iy$ is a complex variable. The symmetry conditions of
the function $F\left(z\right)$ can reduce memory space to store the
function values. It is sufficient to build the table for the values
of function $F\left(z\right)$ in the first quadrant of the complex
plane, i.e., $\left|x\right|\ge0$ and $\left|y\right|\ge0$.

Interpolation techniques are required to predict a value of a function
at a point inside its domain based upon the known tabulated values.
For a given set of data points $\left(z_{i},f_{i}\right)$, $i=0,\dots,N$,
where no two $z_{i}$'s are the same, the interpolated value $g\left(z\right)$
at a value $z\ne z_{i}$ is found from \begin{equation}
g\left(z\right)=\sum_{i=0}^{N}f_{i}L_{i}\left(z\right)\label{eq:lt-2}\end{equation}
 where the $L_{i}$ is Lagrange's $N$-th order polynomials \begin{equation}
L_{i}\left(z\right)=\prod_{j=0,j\ne i}^{N}\frac{z-z_{j}}{z_{i}-z_{j}}.\label{eq:lt-3}\end{equation}
 In order to save the interpolation time further, one can divide $z$-space
and apply a different degree of the Lagrange polynomial. For an example,
we apply a sixth order polynomial for small amplitudes $\left|z\right|\le4\sigma$
while a third order polynomial is applied for $\left|z\right|>4\sigma$,
because the function $F\left(z\right)$ varies more rapidly at small
$\left|z\right|$ and slowly at large $\left|z\right|$ .

\subsubsection{Poisson solver model}

The weak-strong model is a good approximation when one beam has much
smaller intensity than the other, but it is not valid when the intensities
of the two beams are comparable because each beam's parameters are
changed by the other beam. One has to solve for the field of each
beam self-consistently. The fields are the solutions of the Poisson
equation given by \begin{equation}
\nabla^{2}\phi\left(\mathbf{r}\right)=-4\pi\rho\left(\mathbf{r}\right)\label{eq:PS-1}\end{equation}
 where $\phi$ is the electrostatic potential and $\rho$ the density
function of the beam. The solution can be obtained by \begin{equation}
\phi\left(\mathbf{r}\right)=\int G\left(\mathbf{r},\mathbf{r}_{1}\right)\rho\left(\mathbf{r}_{1}\right)d\mathbf{r}_{1}\label{eq:PS-2}\end{equation}
 where $G$ is the Green's function of the Poisson equation and in
two space dimension, is given by \begin{equation}
G\left(x,y:x_{1},y_{1}\right)=-\frac{1}{4\pi}\ln\left[\left(x-x_{1}\right)^{2}+\left(y-y_{1}\right)^{2}\right].\label{eq:PS-3}\end{equation}
 Equation \prettyref{eq:PS-2} can be efficiently calculated using
a convolution theorem and inverse Fourier transform: \begin{equation}
\phi\left(\mathbf{r}\right)=\mathcal{F}^{-1}\left(\hat{G}\left(\boldsymbol{\omega}\right)\hat{\rho}\left(\boldsymbol{\omega}\right)\right)\label{eq:PS-4}\end{equation}
 where $\hat{G}\left(\boldsymbol{\omega}\right)=\left(\frac{1}{\sqrt{2\pi}}\right)^{2}\int_{\mathbb{R}^{2}}G\left(\mathbf{r}\right)e^{-i\boldsymbol{\omega}\cdot\mathbf{r}}d\mathbf{r}$
and $\hat{\rho}\left(\boldsymbol{\omega}\right)=\left(\frac{1}{\sqrt{2\pi}}\right)^{2}\int_{\mathbb{R}^{2}}\rho\left(\mathbf{r}\right)e^{-i\boldsymbol{\omega}\cdot\mathbf{r}}d\mathbf{r}$.
It is assumed in Eq. \prettyref{eq:PS-4} that the density function
$\rho\left(\mathbf{r}\right)$ is periodic in both $x$ and $y$ directions.
However, since the beam has a finite charge distribution surrounded
by a conducting wall in an accelerator system, the transverse beam
density does not meet the periodicity requirement of FFT techniques.
In order to apply the above formalism, the density function should
be rewritten by, in the doubled computational domain \citep{Hockney-1}:
\begin{equation}
\rho_{new}\left(x,y\right)=\begin{cases}
\rho\left(x,y\right) & ,\;0<x\le L_{x},\;0<y\le L_{y}\\
0 & ,\; L_{x}<x\le2L_{x},\; or\; L_{y}<y\le2L_{y}.\end{cases}\label{eq:PS-5}\end{equation}
 Green's function is defined in the doubled domain, as follows: \begin{equation}
G_{new}\left(x,y\right)=\begin{cases}
G\left(x,y\right) & ,\,0<x\le L_{x},\;0<y\le L_{y}\\
G\left(2L_{x}-x,y\right) & ,\, L_{x}<x\le2L_{x},\;0<y\le L_{y}\\
G\left(x,2L_{y}-y\right) & ,\,0<x\le L_{x},\; L_{y}<y\le2L_{y}\\
G\left(2L_{x}-x,2L_{y}-y\right) & ,\, L_{x}<x\le2L_{x},\; L_{y}<y\le2L_{y}.\end{cases}\label{eq:PS-6}\end{equation}
 Both $\rho_{new}$ and $G_{new}$ are doubly periodic functions with
periods $2L_{x}$ and $2L_{y}$. It is noted that only the potential
within a domain $\left(0,L_{x}\right]\times\left(0,L_{y}\right]$
is valid. The potential outside the domain is incorrect, but it doesn't
matter because the physical domain of interest is $\left(0,L_{x}\right]\times\left(0,L_{y}\right]$.
When one beam is separated far from the other, one can apply a shifted
Green's function approach \citep{Qiang}.

\subsubsection{Crossing angle}

\begin{figure}
\begin{centering}
\includegraphics{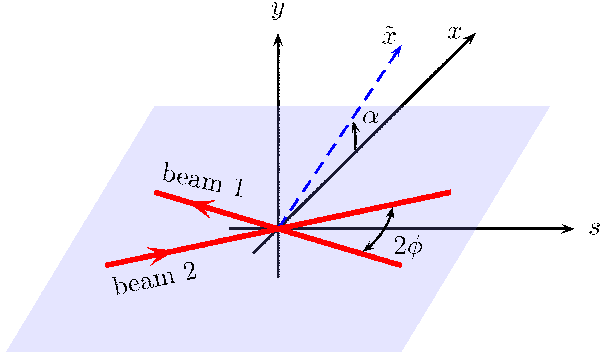} 
\par\end{centering}

\caption{Definition of crossing angles $\alpha$ and $\phi$: $\alpha$ is
the crossing plane angle in the $x-y$ plane and $\phi$ is the half
crossing angle in the $\tilde{x}-s$ plane. $s$ is the axis along
the beam direction when there is no crossing angle. The $\tilde{x}-s$
plane is the crossing plane defined by the angle $\alpha$. The beam
trajectories, shown by lines with arrows, lie in the crossing plane.
\label{fig:xangle}}
\end{figure}

When there exists a finite crossing angle between two colliding beams
at an interaction point, the beam-beam force experienced by a test
particle will have transverse and longitudinal components because
the electric field generated by the opposing beam is not perpendicular
to the particle velocity anymore. The existence of a longitudinal
force makes it difficult to apply the result of previous sections.
A transformation can be used to remedy the difficulty. It transforms
a crossing angle collision in the laboratory frame to a head-on collision
in the rotated and boosted frame which is called the head-on frame
\citep{Hirata-2,Leunissen}. The transformation can be described by
a transformation from the accelerator coordinates to Cartesian coordinates,
a Lorentz boost, and again a backward transformation to the accelerator
coordinates: \begin{equation}
\begin{aligned}x^{*} & =z\cos\alpha\tan\phi+x\left[1+h_{x}^{*}\cos\alpha\sin\phi\right]+yh_{x}^{*}\sin\alpha\sin\phi\\
y^{*} & =z\sin\alpha\tan\phi+y\left[1+h_{y}^{*}\sin\alpha\sin\phi\right]+xh_{y}^{*}\cos\alpha\sin\phi\\
z^{*} & =\frac{z}{\cos\phi}+h_{z}^{*}\left[x\cos\alpha\sin\phi+y\sin\alpha\sin\phi\right]\\
p_{x}^{*} & =\frac{p_{x}}{\cos\phi}-h\cos\alpha\frac{\tan\phi}{\cos\phi}\\
p_{y}^{*} & =\frac{p_{y}}{\cos\phi}-h\sin\alpha\frac{\tan\phi}{\cos\phi}\\
p_{z}^{*} & =p_{z}-p_{x}\cos\alpha\tan\phi-p_{y}\sin\alpha\tan\phi+h\tan^{2}\phi\end{aligned}
\label{eq:ca-1}\end{equation}
 where a star ({*}) stands for a dynamical variable in the head-on
frame, the Hamiltonian $h\left(p_{x},p_{y},p_{z}\right)=p_{z}+1-\sqrt{\left(p_{z}+1\right)^{2}-p_{x}^{2}-p_{y}^{2}}$,
$h_{x}^{*}=\partial h^{*}/\partial p_{x}^{*}$, $h^{*}\left(p_{x}^{*},p_{y}^{*},p_{z}^{*}\right)=h\left(p_{x}^{*},p_{y}^{*},p_{z}^{*}\right)$,
$\alpha$ the crossing plane angle in the $x-y$ plane, and $\phi$
the half crossing angle in the $\tilde{x}-s$ plane as shown in Fig.
\ref{fig:xangle}. %

\subsection{Finite bunch length}

The effects due to the finite (as opposed to infinitesimal) bunch
length need to be considered when the transverse beta functions at
the interaction point are small and comparable to $\sigma_{z}$. The
finite longitudinal length is considered by dividing the beam into
longitudinal slices and by a so called synchro-beam map \citep{Hirata-1}.
We make slices of both beams moving in opposite directions. Each slice
of the strong bunch is integrated over its length, and has only a
transverse charge distribution at its center. We take into account
the collision between a pair of slices: the $i^{th}$ slice of a bunch
and the $j^{th}$ slice of a bunch in the other beam. The collision
takes place at collision point $S\left(z^{i},z_{*}^{j}\right)=\frac{1}{2}\left(z^{i}-z_{*}^{j}\right)$
which is usually different from the interaction point. For example,
the $i^{th}$ slice of a bunch has successive collisions with slices
of a bunch in the other beam. In addition, the electric field varies
along the bunch due to the inhomogeneity of the charge density in
the longitudinal direction, and couples transverse and longitudinal
motions. The coupling can be modeled by the synchro-beam map which
includes beam-beam interactions due to the longitudinal component
of the electric field as well as the transverse components. The transformation
is given by \citep{Hirata-1} \begin{equation}
\begin{aligned}x^{new} & =x+S\left(z,z_{*}\right)\left.\frac{\partial U}{\partial x}\right|_{S},\enskip p_{x}^{new}=p_{x}-\left.\frac{\partial U}{\partial x}\right|_{S}\\
y^{new} & =y+S\left(z,z_{*}\right)\left.\frac{\partial U}{\partial y}\right|_{S},\; p_{y}^{new}=p_{y}-\left.\frac{\partial U}{\partial y}\right|_{S}\\
z^{new} & =z,\:\delta^{new}=\delta-\frac{1}{2}\left.\frac{\partial U}{\partial x}\right|_{S}\left[p_{x}-\frac{1}{2}\left.\frac{\partial U}{\partial x}\right|_{S}\right]-\frac{1}{2}\left.\frac{\partial U}{\partial y}\right|_{S}\left[p_{y}-\frac{1}{2}\left.\frac{\partial U}{\partial y}\right|_{S}\right]-\frac{1}{2}\left.\frac{\partial U}{\partial z}\right|_{S}.\end{aligned}
\label{eq:fl-1}\end{equation}
 Here, $\left.\right|_{S}$ represents the evaluation at the collision
point $S\left(z,z_{*}\right)$. $U$ is the normalized potential energy
$U=q\Phi/E_{0}$ and is given by \begin{equation}
U\left(x,y;\sigma_{x}\left(s\right),\sigma_{y}\left(s\right)\right)=\frac{N_{*}r_{0}}{\gamma}\int_{0}^{\infty}d\zeta\frac{-1+\exp\left(-\frac{x^{2}}{2\sigma_{x}^{2}+\zeta}-\frac{y^{2}}{2\sigma_{y}^{2}+\zeta}\right)}{\sqrt{\left(2\sigma_{x}^{2}+\zeta\right)\left(2\sigma_{y}^{2}+\zeta\right)}}.\label{eq:fl-2}\end{equation}
 The dependence on the bunch length is contained in $\sigma_{x}(s),\sigma_{y}(s)$.
The transverse derivatives of the potential energy are \begin{equation}
\left.\frac{\partial U}{\partial x}\right|_{S}=-\Delta x^{\prime}\left(X,Y;S\left(z,z_{*}\right)\right),\;\left.\frac{\partial U}{\partial y}\right|_{S}=-\Delta y^{\prime}\left(X,Y;S\left(z,z_{*}\right)\right)\label{eq:fl-3}\end{equation}
 where $\left(X,Y\right)$ are the transverse coordinates at $S\left(z,z_{*}\right)$,
and$\Delta x^{\prime}$ and $\Delta y^{\prime}$ are given by Eq.
\prettyref{eq:bb-2}. The longitudinal derivative of the potential
energy which is related to the longitudinal beam-beam kicks is expressed
by \begin{equation}
\begin{aligned}\left.\frac{\partial U}{\partial z}\right|_{S} & =\frac{1}{2}\left.\frac{d\sigma_{x}^{2}}{ds}\frac{\partial U}{\partial\sigma_{x}^{2}}\right|_{s=S\left(z,z_{*}\right)}+\frac{1}{2}\left.\frac{d\sigma_{y}^{2}}{ds}\frac{\partial U}{\partial\sigma_{y}^{2}}\right|_{s=S\left(z,z_{*}\right)}\\
\frac{\partial U}{\partial\sigma_{x}^{2}} & =\frac{1}{2\left(\sigma_{x}^{2}-\sigma_{y}^{2}\right)}\left[x\Delta x^{\prime}+y\Delta y^{\prime}+\frac{2N_{*}r_{0}}{\gamma}\left(\frac{\sigma_{y}}{\sigma_{x}}e^{-\frac{x^{2}}{2\sigma_{x}^{2}}-\frac{y^{2}}{2\sigma_{y}^{2}}}-1\right)\right]\\
\frac{\partial U}{\partial\sigma_{y}^{2}} & =\frac{-1}{2\left(\sigma_{x}^{2}-\sigma_{y}^{2}\right)}\left[x\Delta x^{\prime}+y\Delta y^{\prime}+\frac{2N_{*}r_{0}}{\gamma}\left(\frac{\sigma_{x}}{\sigma_{y}}e^{-\frac{x^{2}}{2\sigma_{x}^{2}}-\frac{y^{2}}{2\sigma_{y}^{2}}}-1\right)\right].\end{aligned}
\label{eq:fl-4}\end{equation}
 Note that $\frac{d\sigma_{x}^{2}}{ds}$ and $\frac{d\sigma_{y}^{2}}{ds}$
have zero amplitude and change their sign at the interaction point
if $\alpha_{x}=\alpha_{y}=0$. Test particles experience longitudinal
acceleration and deceleration passing through the bunch moving in
the opposite direction.

\subsection{Compensation schemes}

In storage-ring colliders, a beam experiences periodic perturbations
when it meets the counter-rotating beam in a common beam pipe. The
head-on beam-beam interactions occur when the beams collide in the
detectors while the long-range interactions occur when the beams are
simultaneously present at the same location but are separated transversely.
The nonlinear forces due to these beam-beam interactions result in
a tune spread and can cause emittance growth, a reduction of beam
life time, and therefore reduce the collider luminosity. The combination
of beam-beam and machine nonlinearities excites betatron resonances
which can cause particles to diffuse into the tails of the beam distribution
and even to the physical aperture. Different compensation methods
have been proposed: a current-carrying wire for the effects of the
long-range interactions \citep{Koutchouk} and an electron lens for
the head-on interactions in proton machines \citep{Tsyganov,Shiltsev-4,Shiltsev-3}.
Beam collisions with a crossing angle at the interaction point are
often necessary in colliders to reduce the effects of the long-range
interactions. The crossing angle reduces the geometrical overlap of
the beams and hence the luminosity. A deflecting mode cavity, also
known as a crab cavity, offers a promising way to compensate the crossing
angle and to realize effective head-on collisions \citep{Palmer,Oide}.
We now describe the modeling of these compensation schemes in the
program.

\subsubsection{Current-carrying wire}

When the separations at long-range interactions are large compared
to the rms beam size the strength of these interactions is inversely
proportional to the distance. Its effect on a beam can be compensated
by a current-carrying wire which creates a magnetic field with the
same $\frac{1}{r}$ dependence. This approach is simple and it is
possible to deal with all multipole orders at once. For a finite length
$l_{w}$ embedded in the middle of a drift length $L$, the transfer
map of a wire can be obtained by \begin{equation}
\mathcal{M}_{w}^{\left(L\right)}=D_{L/2}\circ\mathcal{M}_{k}^{\left(L\right)}\circ D_{L/2}\label{eq:wc-1}\end{equation}
 where $D_{L/2}$ is the drift map with a length $\frac{L}{2}$, and
$\mathcal{M}_{k}^{\left(L\right)}$ is the wire kick integrated over
a drift length. This kick map $\mathcal{M}_{k}^{\left(L\right)}$
is reproduced by the following changes in slope \citep{Erdelyi} \begin{equation}
\left(\begin{array}{c}
\Delta x^{\prime}\\
\Delta y^{\prime}\end{array}\right)=\frac{\mu_{0}}{4\pi}\frac{I_{w}l_{w}}{\left(B\rho\right)}\frac{u-v}{x^{2}+y^{2}}\left(\begin{array}{c}
x\\
y\end{array}\right)\label{eq:wc-2}\end{equation}
 where $I_{w}$ is the current of the wire , $u=\sqrt{\left(\frac{L}{2}+l_{w}\right)^{2}+x^{2}+y^{2}}$
and $v=\sqrt{\left(\frac{L}{2}-l_{w}\right)^{2}+x^{2}+y^{2}}$. We
also take into account the wire misalignment including pitch and yaw
angles $\left(\theta_{x},\theta_{y}\right)$ respectively as well
as lateral shifts $\left(\Delta x,\Delta y\right)$. The transfer
map of a wire can be written as \begin{equation}
\mathcal{M}_{w}=S_{\Delta x,\Delta y}\circ T_{\theta_{x},\theta_{y}}^{-1}\circ D_{L/2}\circ\mathcal{M}_{k}^{\left(L\right)}\circ D_{L/2}\circ T_{\theta_{x},\theta_{y}}\label{eq:wc-3}\end{equation}
 where $T_{\theta_{x},\theta_{y}}$ represents the tilt of the coordinate
system by horizontal and vertical angles $\theta_{x},\theta_{y}$
to orient the coordinate system parallel to the wire, and $S_{\Delta x,\Delta y}$
represents a shift of the coordinate axes to make the coordinate systems
after and before the wire agree. When the wire is parallel to the
beam, Eq. \prettyref{eq:wc-3} becomes Eq. \prettyref{eq:wc-1}. For
canceling the long-range beam-beam interactions of the round beam
with the wire, one can get the desired wire current and length by
equating Eq. \prettyref{eq:wc-2} and Eq. \prettyref{eq:bb-2}; the
integrated strength of the wire compensator is related to the integrated
current of the beam bunch as $I_{w}l_{w}=cqN$.

\subsubsection{Electron lens}

For the head-on proton-proton beam collisions, particles of one proton
bunch are focused by a space charge of the counter-rotating proton
bunch. The beam-beam effect on the particles of the proton bunch can
be compensated by a counter-rotating beam of negatively charged particles,
for example, a low-energy electron beam. In order to cancel out the
transverse kick by the counter-rotating proton bunch, the electron
beam should have the same transverse charge profile and current as
the proton bunch. The proton bunch typically exhibits an approximately
Gaussian transverse profile. If we choose a Gaussian distribution
of the electron beam, the transverse kick on particles of the proton
bunch from the electron beam is given by \begin{equation}
\left(\begin{array}{c}
\Delta x^{\prime}\\
\Delta y^{\prime}\end{array}\right)=-\frac{2N_{e}r_{0}}{\gamma r^{2}}\zeta\left(x,y:\sigma_{e}\right)\left(\begin{array}{c}
x\\
y\end{array}\right)\label{eq:el-1}\end{equation}
 where $N_{e}$ is the number of electrons of the electron beam adjusted
by the electron beam speed, $r_{0}$ the classic proton radius, $\gamma$
the Lorentz factor, $r^{2}=x^{2}+y^{2}$, and $\sigma_{e}$ the transverse
beam size of the electron beam. The function $\zeta$ is given by
\begin{equation}
\zeta\left(x,y:\sigma_{e}\right)=\left[1-\exp\left(-\frac{x^{2}+y^{2}}{2\sigma_{e}}\right)\right].\label{eq:el-2}\end{equation}
 For a non-Gaussian electron charge distribution we implement a flat
top profile with smooth edges that generates a linear beam-beam force
near the beam center. This flat top beam profile $\rho_{e}\left(r\right)=\rho_{0}/\left(1+\left(r/\sigma_{e}\right)^{8}\right)$
delivers the transverse kicks given by Eq. \prettyref{eq:el-1}, but
the function $\zeta$ is as follows: \begin{equation}
\zeta=\frac{\sqrt{2}\tilde{\rho}_{0}}{8}\left[\frac{1}{2}\log\left(\frac{\theta_{+}^{2}+1}{\theta_{-}^{2}+1}\right)+\tan^{-1}\theta_{+}+\tan^{-1}\theta_{-}\right]\label{eq:el-3}\end{equation}
 where $\tilde{\rho}$ is a constant, and $\theta_{\pm}=\sqrt{2}\left(\frac{r}{\sigma_{e}}\right)^{2}\pm1$.

\subsubsection{Crab cavity}

When a particle passes through a crab cavity structure, it experiences
a transverse deflection and a small change in its longitudinal energy.
Crab cavities can compensate for the horizontal or vertical crossing
angle at the interaction point by delivering oppositely directed transverse
kicks to the head and the tail of the bunches. In the case of a horizontal
crossing, the kicks from the crab cavity are given by \begin{equation}
\Delta x^{\prime}=-\frac{qV}{E_{0}}\sin\left(\phi_{s}+\frac{\omega z}{c}\right),\enskip\Delta\delta=-\frac{qV}{E_{0}}\cos\left(\phi_{s}+\frac{\omega z}{c}\right)\cdot\frac{\omega}{c}x\label{eq:cc-1}\end{equation}
 where $q$ denotes the particle charge, $V$ the voltage of crab
cavity, $E_{0}$ the particle energy, $\phi_{s}$ the phase of the
synchronous particle with respect to the crab-cavity rf wave, $\omega$
the angular frequency of the crab cavity, $c$ the speed of light,
$z$ the longitudinal coordinate of the particle with respect to the
bunch center, and $x$ the horizontal coordinate. In general this
is a nonlinear map which introduces synchro-betatron coupling, but
for small $z$, this reduces to a linear map in the horizontal-longitudinal
plane. The crab cavity causes a closed orbit distortion dependent
on the longitudinal position of particles, and the beam envelope is
tilted all around the ring. For a bunch shorter than the rf wavelength
of the crab cavity deflecting mode, the tilt angle of the beam envelope
at a location with a beam position monitor (BPM) is given by \begin{equation}
\tan\theta_{crab}=\frac{qV\omega\sqrt{\beta\beta_{crab}}}{c^{2}p_{0}}\left|\frac{\cos\left(\Delta\varphi-\pi Q\right)}{2\sin\pi Q}\right|\label{eq:cc-2}\end{equation}
 where $\beta$ is the beta function at the BPM position, $\beta_{crab}$
the beta function at the crab cavity, $\Delta\varphi$ the phase advance
between the crab cavity location and the BPM, and $Q$ the betatron
tune. The simulations of a crab cavity in the SPS accelerator at CERN
using \texttt{BBSIM} will be described in another paper.

\subsection{Particle distribution}

At the beginning of a simulation, the simulation particles are distributed
over the phase space $\mathbf{x}=\left(x,x^{\prime},y,y^{\prime},z,\delta\right)^{T}$,
called the initial loading. In any simulation the number of particles
$N$ is limited by the computational power. In order to make the best
use of a small number of simulation particles compared to the real
number of particles in the accelerator, the loading should be optimized.
Indeed the initial loading is very important because this choice can
reduce the statistical noise in the physical quantities.

\emph{Gaussian distribution}: For long-term particle tracking where
we calculate emittance growth, we consider an exponential distribution
in action (Gaussian distribution in coordinates) of the form: \begin{equation}
\rho\left(\mathbf{x}\right)=\rho_{0}\exp\left(-\frac{J_{x}}{2\sigma_{J_{x}}}-\frac{J_{y}}{2\sigma_{J_{y}}}-\frac{J_{z}}{2\sigma_{J_{z}}}\right)\label{eq:dt-1}\end{equation}
 where $J_{x}$, $J_{y}$, and $J_{z}$ are the transverse and longitudinal
action variables defined by \begin{equation}
\begin{aligned}J_{x} & =\frac{1}{2\beta_{x}}\left[x^{2}+\left(\beta_{x}x^{'}+\alpha_{x}x\right)^{2}\right],\; J_{y}=\frac{1}{2\beta_{y}}\left[y^{2}+\left(\beta_{y}y^{'}+\alpha_{y}y\right)^{2}\right]\\
J_{z} & =\frac{8}{\pi}\frac{R\nu_{s}}{h^{2}\left|\eta\right|}\left[E\left(k\right)-\left(1-k^{2}\right)K\left(k\right)\right]\end{aligned}
\label{eq:pd-2}\end{equation}
 where $R$ is the radius of the accelerator, $h$ the harmonic number,
$\nu_{s}$ the longitudinal tune, $E$ and $K$ the complete elliptical
integrals, and \begin{equation}
k^{2}=\frac{1}{4}\frac{h^{2}\eta^{2}}{\nu_{s}^{2}}\left(\frac{\Delta p}{p}\right)^{2}+\sin^{2}\frac{\phi}{2}.\label{eq:pd-3}\end{equation}

$\sigma_{J_{x}}$, $\sigma_{J_{y}}$, and $\sigma_{J_{z}}$ are the
rms sizes of action variables. The simulation particles are generated
by two steps: 
\begin{enumerate}
\item The action variables $\left(J_{x},J_{y},J_{z}\right)$ of particles
can be directly generated from the distribution function by the inverse
transform method and the bit-reversed sequence \citep{Birdsall}. 
\item For example, $x$ and $x^{\prime}$ are correlated and their distribution
is $\hat{\rho}\left(x,x^{\prime}\right)=\hat{\rho}_{0}\exp\left(-\frac{x^{2}+\left(\beta_{x}x^{\prime}+\alpha_{x}x\right)^{2}}{2\sigma_{x}^{2}}\right)$.
Since the horizontal action $J_{x}$ is determined at the first step,
the horizontal coordinates $\left(x,x^{\prime}\right)$ can be obtained
from the random variates: \[
x=\sqrt{J_{x}}\cos\theta_{x},\quad x^{\prime}=\sqrt{J_{x}}\left(\sin\theta_{x}-\alpha_{x}\cos\theta_{x}\right)/\beta_{x}\]
 where the value of $\theta_{x}$ is randomly distributed within the
interval $0\le\theta_{x}\le2\pi$. 
\end{enumerate}
\emph{\indent Hollow Gaussian distribution}: In most cases of particle
tracking, lost particles are observed only above a certain large transverse
action while the beam core is stable. An example is shown in Section
\ref{sub:Tevatron}. A hollow beam is a beam with zero central intensity
along the longitudinal beam axis. For the generation of a hollow beam,
a bunched beam distribution in longitudinal phase space is a Gaussian,
but a distribution in transverse phase space is a hollow Gaussian.
The procedure of generating the hollow distribution is the same as
that for the Gaussian distribution except that the amplitude of transverse
action of a particle should be larger than a minimum value, i.e.,
$J_{x}+J_{y}\ge\sigma_{J}$. Since most of the stable particles are
not included in the tracking simulation, the hollow beam model simulates
a large transverse amplitude Gaussian distribution using a small number
of macro-particles. This distribution is useful when calculating beam
lifetimes.

\subsection{Particle diffusion}

Diffusion coefficients can characterize the effects of the nonlinearities
present in an accelerator, and can be used to find numerical solutions
of a diffusion equation for the density \citep{tsen-1,hjkim-1}. The
solutions yield the time evolution of the beam density distribution
function for a given set of machine and beam parameters. This technique
enables us to follow the beam intensity and emittance growth for the
duration of a luminosity store, something that is not feasible with
direct particle tracking. The transverse diffusion coefficients can
be calculated numerically from \begin{equation}
\begin{aligned}D_{ij}\left(a_{i},a_{j}\right) & =\frac{1}{N}\left\langle \left(J_{i}(a_{i},N)-J_{i}(a_{i},0)\right)\left(J_{j}(a_{j},N)-J_{j}(a_{j},0)\right)\right\rangle \end{aligned}
\label{eq:pd-1}\end{equation}
 where $J_{i}\left(a_{i},0\right)$ is the initial action at an amplitude
$a_{i}$, $J_{i}\left(a_{i},N\right)$ the action with initial amplitude
$a_{i}$ after $N$ turns, $\left\langle \right\rangle $ the average
over simulation particles, and $(i,j)$ are the horizontal $x$ or
the vertical $y$ coordinates. Equation \prettyref{eq:dt-1} is averaged
over a certain number of turns to eliminate the fluctuation in action
due to the phase space structure, e.g. resonance islands. These diffusion
coefficients can be directly used to compare amplitude growth under
different circumstances, e.g with different tunes. Emittance growth
and beam lifetimes can be calculated when these coefficients are used
in a diffusion equation, as mentioned above.

\subsection{Symplecticity \label{sub:Symplecticity}}

In the absence of dissipative effects, particle motion in an accelerator
can be described by Hamilton's equations of motion. Hamiltonian systems
obey the symplectic condition which guarantees the conservation of
phase space volume as the system evolves, this is also known as Liouville's
theorem. For transfer maps described in previous subsections the symplectic
condition requires \begin{equation}
M^{T}SM=S,\quad S=\left(\begin{array}{ccc}
s & 0 & 0\\
0 & s & 0\\
0 & 0 & s\end{array}\right)\label{eq:sm-1}\end{equation}
 where $s=\left(\begin{array}{cc}
0 & 1\\
-1 & 0\end{array}\right)$ is an antisymmetric $2\times2$ matrix, and $M$ is a transfer matrix
for a linear system or the Jacobian matrix of a nonlinear map around
any particle trajectory. For a nonlinear map $\mathcal{M}:\mathbf{x}\longrightarrow\bar{\mathbf{x}}$,
the Jacobian matrix is obtained from first-order partial derivatives
of the new coordinates with respect to the old ones. The elements
are defined as $M_{ij}=\partial\bar{x}_{i}/\partial x_{j}$. During
implementation of the maps for beam dynamics, one should check to
ensure that the map is as symplectic as possible. As a measure of
the symplecticity, a matrix norm of $\left\Vert M^{T}SM-S\right\Vert $
is used in \texttt{BBSIM}. The accuracy of the look-up table model
mentioned in subsection \ref{sub:lookup-table}, for example, depends
on the number of sample points in a given complex space needed for
interpolating the function. Poor interpolation accuracy may violate
the symplecticity, and lead to emittance blow-up or shrinkage. We
use the symplectic norm obtained with the direct calculation of the
complex error function as the benchmark. We find for example, that
in order to maintain the symplectic norm with 70 long-range beam-beam
interactions in the Tevatron, the number of sample points should be
more than 4 points per rms beam size.

\subsection{Diagnostics}

Numerical simulation enables the generation of very large amounts
of data. The \texttt{BBSIM} code monitors physical quantities, for
example, particle amplitudes and saves them into an external file
during the simulation. According to a problem of interest, the quantities
to be saved can be chosen in order to extract valuable information
from post-processing. In addition, some diagnostic functions are calculated
in the code as follows:

\emph{Betatron tune distribution}: The betatron tune in an accelerator
is one of the most important beam parameters. The tune of each particle
in the beam distribution is calculated with a Hanning filter applied
to an fast-Fourier transform of particle coordinates found from tracking
\citep{Bartolini}.

\emph{Beam transfer function}: The beam transfer function (BTF) is
defined as the beam response to a small external longitudinal or transverse
excitation at a given frequency. BTF diagnostics are widely employed
in accelerators due to its non-destructive nature. A stripline kicker
or rf cavity excites betatron or synchrotron oscillations respectively
over the appropriate tune spectrum. The beam response is observed
in a downstream pickup. The fundamental applications of BTF are to
measure the transverse tune and tune distribution by exciting betatron
oscillation, to analyze the beam stability limits, and to determine
the impedance characteristics of the chamber wall, and feedback system
\citep{Borer}. In the code, we apply a sinusoidal driving force to
a beam in a transverse plane. The driving frequency is swept in equidistant
steps over a continuous frequency range which includes the betatron
tune. At each new frequency there is initially a transient response
which must be allowed to relax before the frequency is extracted from
the data. We avoid the issue of the transients in the simulations
by reloading the initial particle distribution at each new frequency.

\emph{Frequency diffusion}: We have calculated frequency diffusion
maps as another way to investigate the effects of nonlinear forces.
The map represents the variation of the betatron tunes over two successive
sets of the tunes \citep{Laskar}: The variation can be quantified
by $d=\log\sqrt{\Delta\nu_{x}^{2}+\Delta\nu_{y}^{2}}$, where ($\Delta\nu_{x}=\nu_{x}^{\left(2\right)}-\nu_{x}^{\left(1\right)},\Delta\nu_{y}=\nu_{y}^{\left(2\right)}-\nu_{y}^{\left(1\right)}$)
are the tune variations between the first set and next set of 1024
turns. If the tunes $\left(\nu_{x}^{\left(1\right)},\nu_{y}^{\left(1\right)}\right)$
are different from $\left(\nu_{x}^{\left(2\right)},\nu_{y}^{\left(2\right)}\right)$,
the particle is moving to different amplitudes. A large tune variation
is generally an indicator of fast diffusion and reduced stability.

\emph{Dynamic aperture}: The dynamic aperture of an accelerator is
defined as the smallest radial amplitude of particles that survive
up to a certain time interval, for example, $10^{6}$ turns. As the
number of turns increases, the dynamic aperture approaches an asymptotic
value. Initial particles are distributed uniformly over the transverse
phase space with amplitudes typically varying between 0-20 $\sigma$,
where $\sigma$ is the rms transverse beam size. The longitudinal
amplitude is chosen as largest value within a bunch.

\emph{Emittance}: The emittance is defined as the area (or volume)
of phase space enclosed by the ellipse containing all the particles
in its interior. Statistically, the rms beam emittance can be calculated
by a determinant of $\Sigma$-matrix of a beam distribution: \begin{equation}
\epsilon=\left[\det\left(\Sigma\right)\right]^{1/d}\label{eq:DE-1}\end{equation}
 where $d$ is the dimension of phase space, the element of $\Sigma$-matrix
is $\Sigma_{ij}=\left\langle \left(\zeta_{i}-\left\langle \zeta_{i}\right\rangle \right)\left(\zeta_{j}-\left\langle \zeta_{j}\right\rangle \right)\right\rangle $,
and $\zeta=\left\{ x,x^{\prime},y,y^{\prime},z,\delta\right\} $.
For example, horizontal emittance is obtained by $\epsilon_{x}=\left[\det\left(\begin{array}{cc}
\Sigma_{xx} & \Sigma_{xx^{\prime}}\\
\Sigma_{x^{\prime}x} & \Sigma_{x^{\prime}x^{\prime}}\end{array}\right)\right]^{1/2}$. In addition to the emittance of each degree of freedom, four- and
six-dimensional emittances are calculated to see the correlation and
coupling between the phase space coordinates.

\emph{Beam loss}: The beam loss is one of the fundamental observables
and it can be directly compared with simulation. During a beam simulation,
each particle is monitored if it reaches a physical aperture transversely
or the rf bucket longitudinally. The particle passing beyond the aperture
is considered as a lost particle. Unlike a real machine, several \emph{virtual}
apertures are placed inside a beam pipe. The multiple apertures are
used to find beam losses at different apertures.

\section{Parallelization \label{sec:Parallelization}}

Realistic simulations of beam dynamics demand large computational
resources. Calculations on these large number of particles can be
distributed over several processors of a parallel computer to improve
performance. Two basic approaches exist to allocate the calculations
to the processors, particle based and domain (space) based partitions.
In the former approach, the particles are uniformly allocated to the
processors. They are not limited to a certain spatial domain. The
completion time of a parallel solution depends on the processor with
the maximum computational workload. The particle decomposition can
distribute the computational load evenly among all processors while
the interaction between particles, for example, intra-beam scattering
needs a very large number of communications between processors since
the interacting particles can be located in a distant processor. Conversely,
in the domain decomposition approach, the spatial domain is partitioned
into elementary regions, and each processor is responsible for one
of these regions. The particles in the accelerator simulation are
transported by the lattice map. The map causes significant particle
movement which may cause the load to become quickly unbalanced. The
simulation of colliding beams has two aspects, i.e., pure particle
transport and electromagnetic field evaluation. The domain deposition
approach is an efficient way of parallelizing the field solver. To
achieve the workload balanced, our approach is to use both decomposition
schemes.

\begin{figure}
\begin{centering}
\includegraphics[scale=0.55]{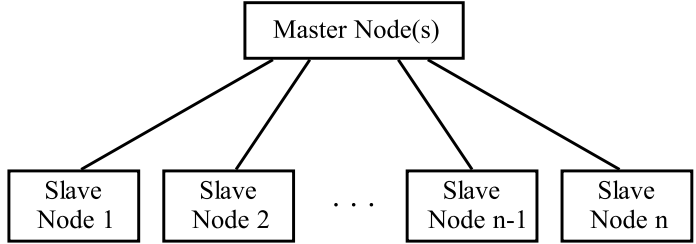} 
\par\end{centering}

\caption{Master/slave communication diagram. \label{fig:comm-diag}}
\end{figure}

We have implemented a parallel calculation in the \texttt{BBSIM} code
to perform a tracking simulation of large numbers of particles. When
the weak-strong beam-beam model is used, only the particle decomposition
scheme can be applied for parallel computation. Its implementation
can be made trivially because the macroparticles are never moved from
one processor to another. No inter-processor communication is necessary
while the particle trajectories are being developed. Most calculations
on each node are executed sequentially. In this model the communication
between the parallel processes is only required for reading input
data, generating an initial beam distribution, calculating diagnostics
such as beam emittance, and writing out the diagnostic information.
For the Poisson solver model, however, we have used a particle-in-cell
(PIC) model to update the electromagnetic field. The PIC model represents
the beam as a large number of computational particles moving according
to classical mechanics. The PIC algorithm can be characterized as
follows: (a) integrate over particles to obtain a charge distribution
on the grid point, (b) solve a Poisson equation for the potential,
and (c) interpolate the potential or field onto particles for a small
interval of time to advance the position and velocity of particles.
Part (a) requires $\mathcal{O}\left(N_{g}^{d}\right)$ numeric operations
for a FFT Poisson solver, where $N_{g}$ is the number of grid points
per dimension and $d$ is the number of degrees of freedom. Part (a)
and (c) obviously require $\mathcal{O}\left(N_{p}\right)$ operations,
where $N_{p}$ is the number of computation particles. In general,
$N_{p}$ is much larger than $N_{g}$ in that the number of particles
should increase according to the degree of freedom to maintain the
statistical noise to be constant in a higher spatial dimension. The
particle calculations thus dominate the overall computational process,
which suggests a prior parallelization of particle calculation. Master/slave
configuration of computational nodes shown in Fig. \ref{fig:comm-diag}
is considered due to the difference of numeric operations between
particles and field updates.

Each processor on the master and slave nodes possesses the same number
of particles. All processors are responsible for advancing their particles.
On the contrary, the master node may be a single or many processor(s),
depending on the number of grid points required. The charge density
of a beam is deposited on the computational grids of each processor
using standard area weighting (or higher order) methods \citep{Hockney-2}.
The master node gathers the charge density from all processors, and
solves the Poisson equations in parallel. The master node broadcasts
the solution of the electric field to all processors such that each
processor exerts the electromagnetic force on the particles owned
by the processor. %

\begin{figure}
\begin{centering}
\subfloat[]{\begin{centering}
\includegraphics[bb=20bp 5bp 340bp 205bp,clip,scale=0.7]{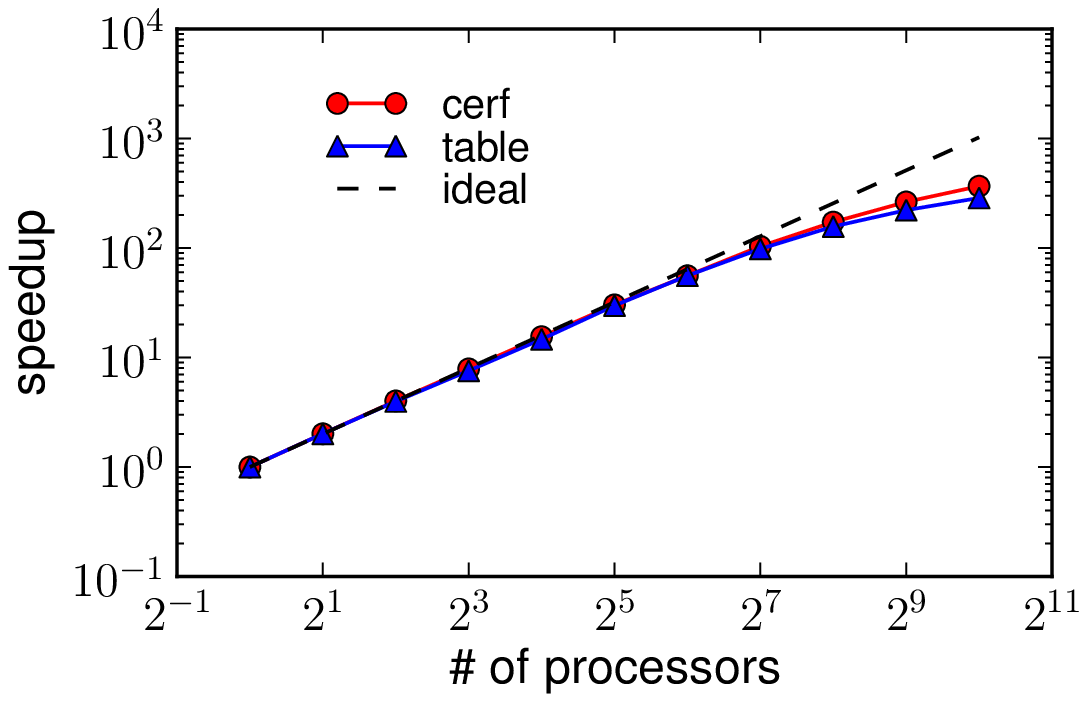}
\par\end{centering}

}\subfloat[]{\begin{centering}
\includegraphics[bb=20bp 5bp 340bp 205bp,clip,scale=0.7]{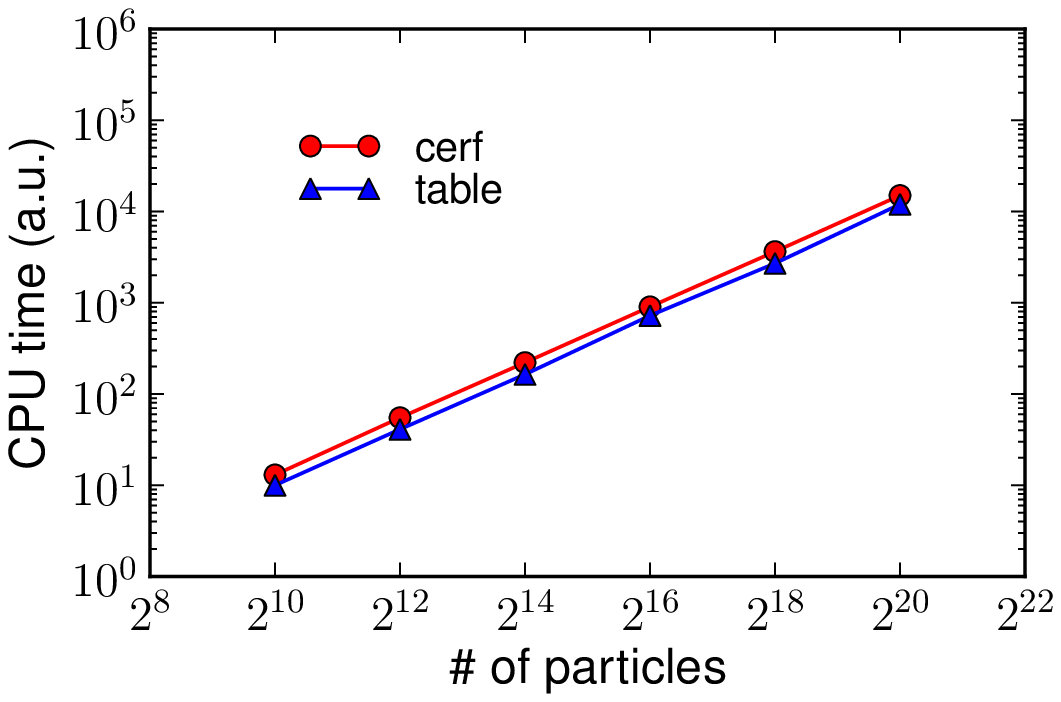}
\par\end{centering}

} 
\par\end{centering}

\caption{Plots of (b) parallel speedup versus the number of nodes, and (b)
CPU time versus the number of simulation particles. cerf and table
represent the weak-strong model, and look-up table model respectively.\label{fig:cpu-time}}
\end{figure}

The performance of the master/slave parallelization approach has been
investigated using a real lattice of the Tevatron which has two head-on
beam-beam collisions and 70 long-range beam-beam interactions. Speedup
test has been performed on the Cray XT5 of the National Energy Research
Scientific Computing Center at Lawrence Berkeley National Laboratory.
The system is built up of 664 nodes with two quad-core AMD 2.4 GHz
processors per node. The speedup of a parallel program is a measure
of the utilization of parallel resources and is simply defined as
the ratio between sequential execution time and parallel execution
time \citep{Quinn}: \begin{equation}
S_{p}=\frac{T_{1}}{T_{p}}\label{eq:pa-1}\end{equation}
 where $p$ is the number of processors, $T_{1}$ is the execution
time of the sequential algorithm, and $T_{p}$ is the execution time
of the parallel algorithm with $p$ processors. For a fixed number
of processors $p$, typically the speedup is $0<S_{p}\le p$. Ideally
all parallel programs should exhibit a linear speedup, i.e., $S_{p}=p$,
but it is not common because communication between processors is considerably
slower than computation in each processor. Figure \ref{fig:cpu-time}
(a) illustrates the resulting speedup as a function of the number
of processors. %

The parallelization speedup based on the total simulation time is
compared for simulations with the weak-strong model and the look-up
table model. The speedup curves are very close to the ideal one below
a certain number of processors, while they are less than optimal when
the number of processors increases above a critical value, for example,
$2^{6}$ processors. On large numbers of processors a relative fraction
of the communication time in the total computing time becomes large.
A parallel efficiency, defined as the speedup factor divided by the
number of processors, can be obtained as high as 87\% up to the critical
number of processors. Though the efficiency falls well below 38\%
when the number of processors is beyond $2^{10}$, it runs 367 times
faster than on a single processor. In order to see the scalability
of our parallel code for larger problem sizes, Fig. \ref{fig:cpu-time}
(b) shows the execution time as a function of the number of macro-particles.
Here the number of processors is fixed at $2^{6}$ for all cases.
It is seen that with increasing the number of simulation particles,
the execution time also increases linearly.

\section{Applications \label{sec:Application}}

In high energy storage-ring colliders, the beam-beam interactions
cause emittance growth, may reduce beam lifetime, and hence limit
the collider luminosity. We have used \texttt{BBSIM} to study beam-beam
interactions and their compensations in the Tevatron, in RHIC and
in the LHC.

\subsection{Tevatron\label{sub:Tevatron}}

\begin{figure}
\begin{centering}
\includegraphics[bb=0bp 0bp 504bp 180bp,clip]{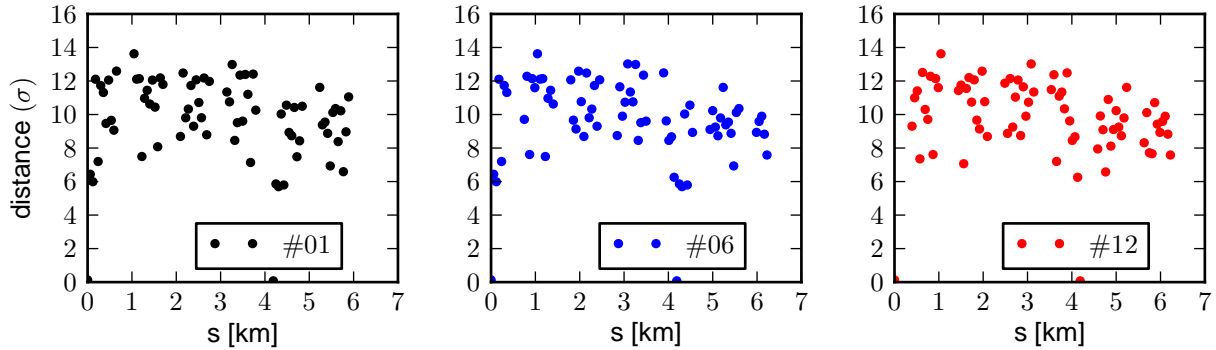} 
\par\end{centering}

\caption{Separation distance between proton and anti-proton beams for anti-proton
bunches \#1, \#6 and \#12. The separation is normalized by proton
beam's rms size. \label{fig:tev-sep}}
\end{figure}

The luminosity of a collider is found from \begin{equation}
\mathcal{L}=\frac{N_{1}N_{2}fN_{B}}{4\pi\sigma_{x}\sigma_{y}}R\label{eq:AT-1}\end{equation}
 where $N_{1}$ and $N_{2}$ are the bunch populations of the colliding
beams, $f$ the revolution frequency, $N_{B}$ the number of bunches
in one beam, $\sigma_{x}$ and $\sigma_{y}$ the horizontal and vertical
rms beam sizes at the collision points respectively, and $R$ the
luminosity reduction factor due to the {}``hour-glass'' effect and
due to non-zero crossing angle at the interaction point. The beam-beam
tune shift of beam 1 is proportional to the factor $N_{2}/\sigma_{x}\sigma_{y}$
and experience from colliders worldwide has shown that the achievable
tune shift (and hence luminosity) is limited by the dynamics of the
beam-beam interaction. In the Tevatron, proton and anti-proton bunches
collide at two detectors called CDF and D0. They share the same beam
pipe. Since the two beams circulate on helical orbits, the optics
and dynamics of the beam-beam interactions are complex. The beam-beam
interactions occur all around the ring and at varying betatron phases.
In run II, each beam has three trains of 12 bunches \citep{tev-runii}.
Each bunch experiences 72 interactions: two interactions are the head-on
collisions in the detectors. However, the other 70 interactions are
long-range, and are placed at different locations for each bunch.
Consequently the beam separation distances between proton and anti-proton
beams at the long-range locations are different from bunch to bunch.
Figure \ref{fig:tev-sep} shows the radial beam separation of three
anti-proton bunches from the proton bunches in units of the rms beam
size of the proton beam at the locations of the beam-beam interactions.
 The long-range interactions of special importance are those on either
side of the head-on interaction points. These occur at small separations
and the beta functions there are large. It was observed that the emittance
growth at the end bunches of each train is smaller than those in the
middle of the train. Here we choose two end bunches (\#1 and \#12)
and one middle bunch (\#6) of the first train.

\begin{figure}
\begin{centering}
\subfloat[]{\begin{centering}
\includegraphics[bb=20bp 5bp 340bp 205bp,clip,scale=0.7]{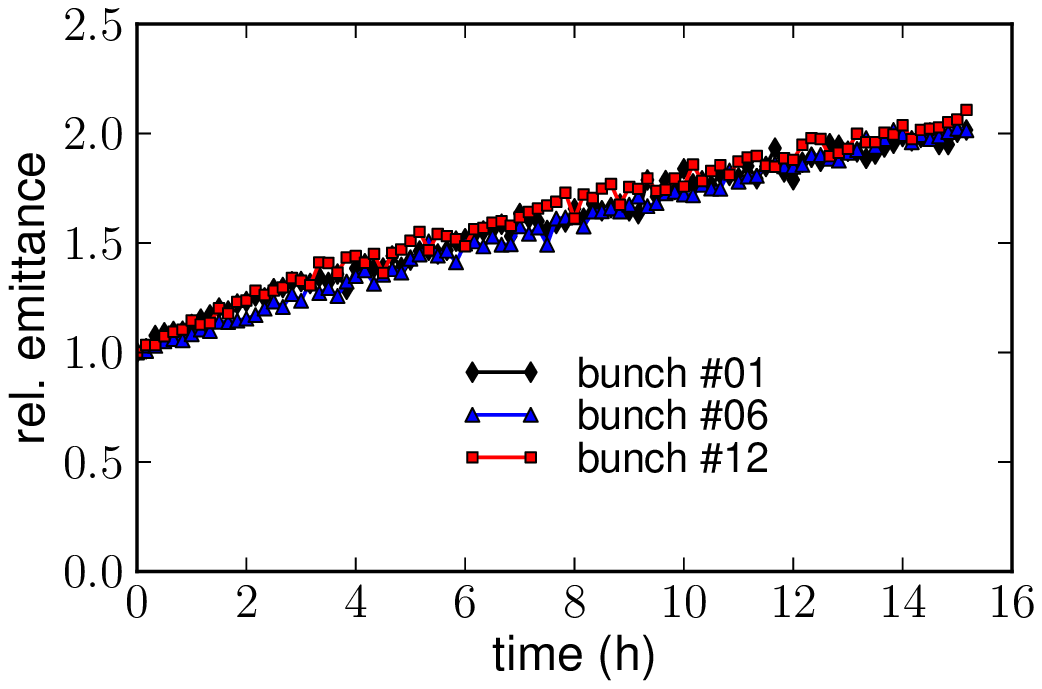}
\par\end{centering}

}\subfloat[]{\begin{centering}
\includegraphics[bb=20bp 5bp 340bp 205bp,clip,scale=0.7]{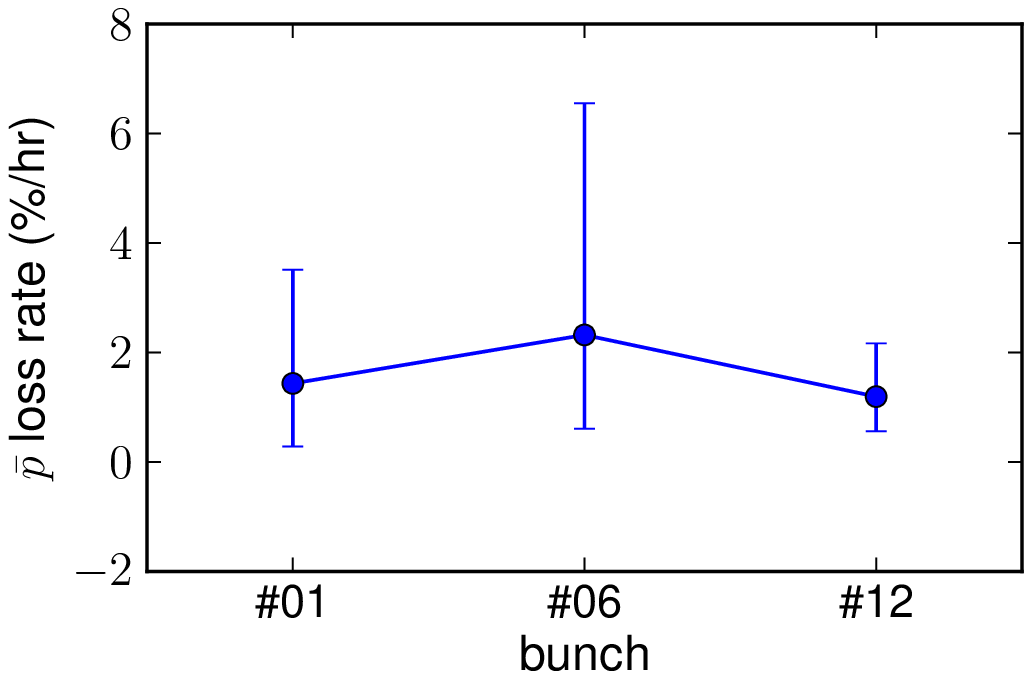}
\par\end{centering}

}
\par\end{centering}

\begin{centering}
\subfloat[]{\begin{centering}
\includegraphics[bb=20bp 5bp 340bp 205bp,clip,scale=0.7]{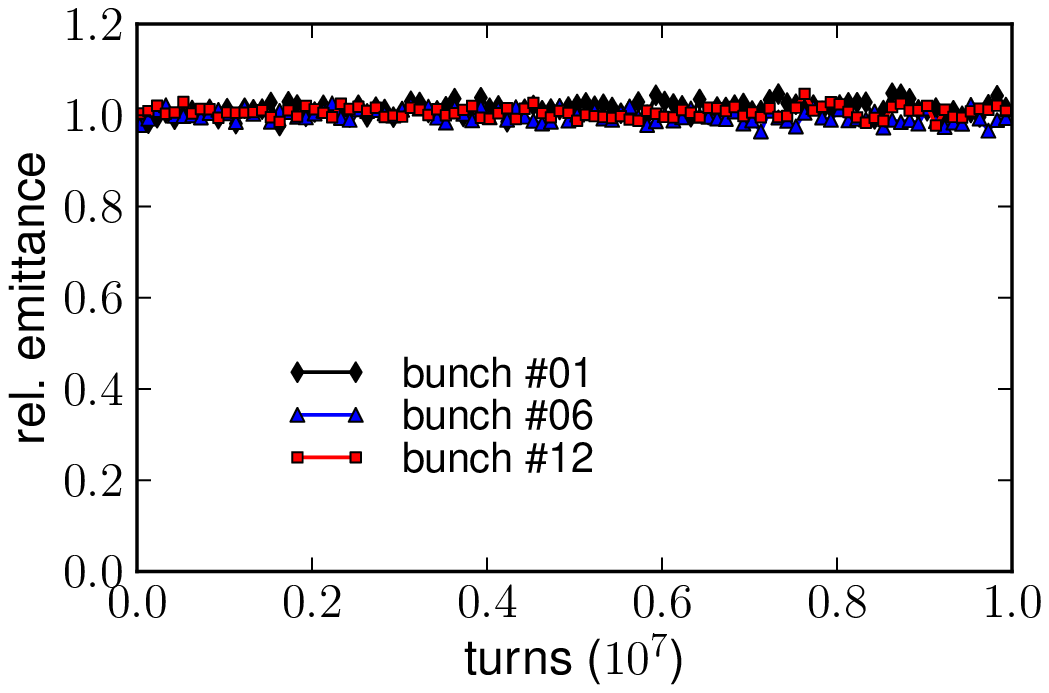}
\par\end{centering}

}\subfloat[]{\begin{centering}
\includegraphics[bb=20bp 5bp 340bp 205bp,clip,scale=0.7]{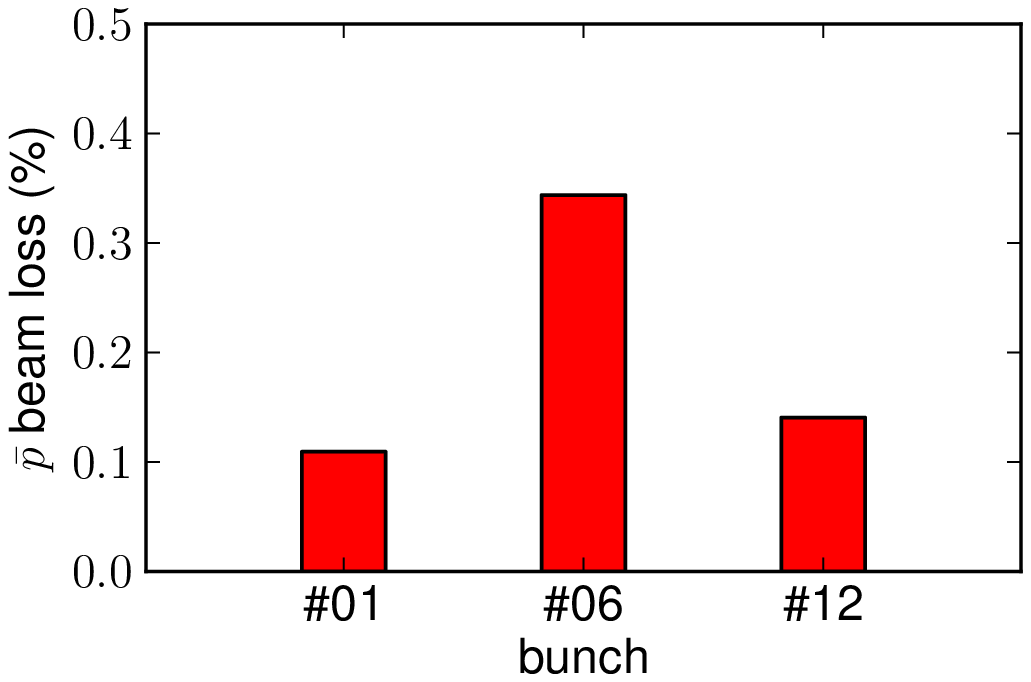}
\par\end{centering}

}
\par\end{centering}

\caption{(a) Variation of anti-proton emittance of three bunches, \#1, \#6,
and \#12, of store \#7650, (b) non-luminous loss rates of anti-proton
during the first 1 hour of stores \#7601-\#7650, (c) simulation of
anti-proton emittance growth, and (d) simulation of anti-proton beam
loss. Here the emittance is plotted as $\epsilon_{4d}=\sqrt{\epsilon_{x}\epsilon_{y}}$.
In the simulation, initial anti-proton emittance $\left(\epsilon_{x},\epsilon_{y}\right)$
is (9.0,7.8) mm-mrad, bunch length 1.5 nsec, and bunch intensity $0.86\times10^{11}$.
Proton's initial emittance is (18,23) mm-mrad, bunch length 1.7 nsec,
bunch intensity $2.64\times10^{11}$. Nominal tune is (20.571, 20.569).
Revolution frequency is 47.7 kHz. \label{fig:tev-loss}}
\end{figure}

\begin{figure}
\begin{centering}
\subfloat[]{\begin{centering}
\includegraphics[bb=20bp 5bp 340bp 280bp,clip,scale=0.6]{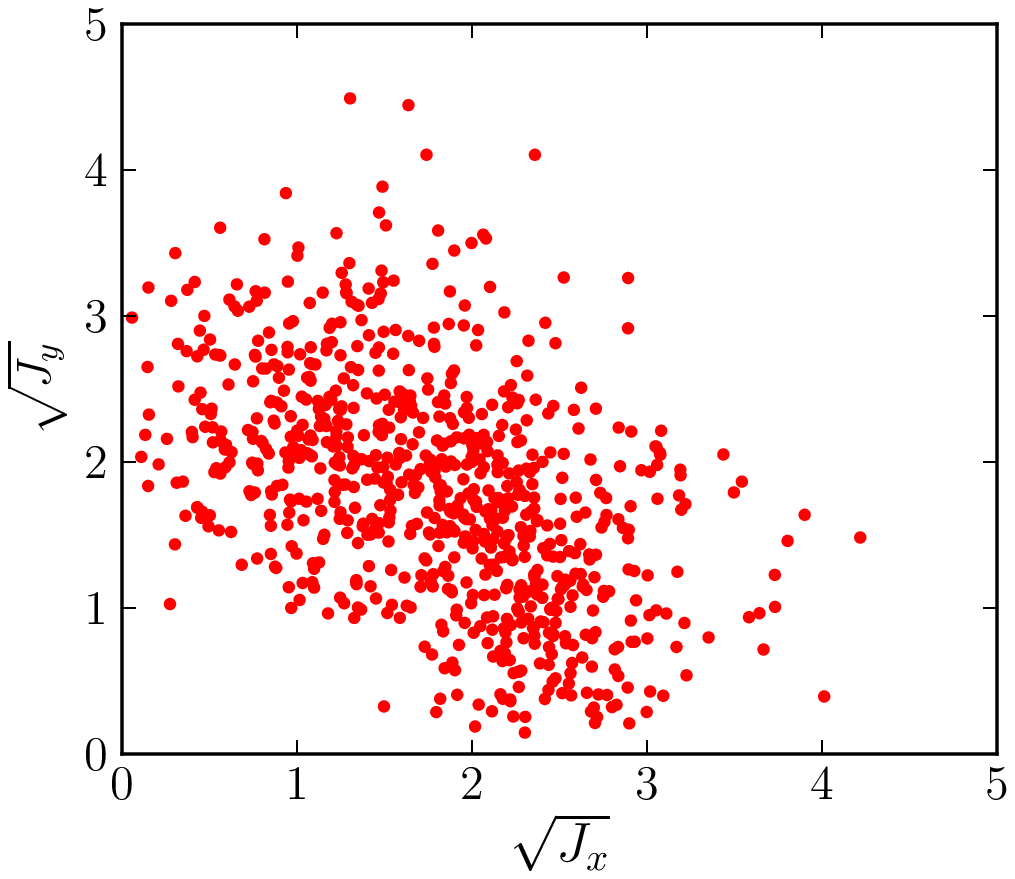}
\par\end{centering}

}\subfloat[]{\begin{centering}
\includegraphics[bb=20bp 5bp 340bp 280bp,clip,scale=0.6]{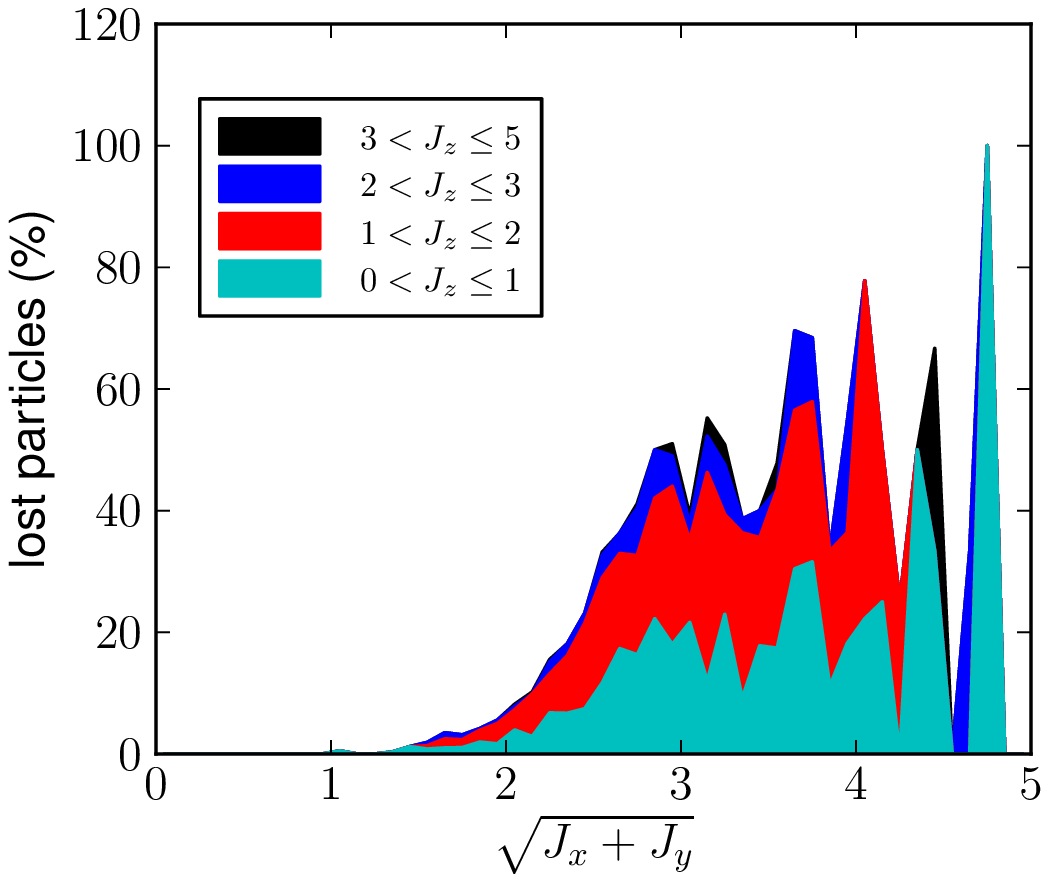}
\par\end{centering}

} 
\par\end{centering}

\caption{(a) Scatter plot of lost particles in action space $\left(\sqrt{J_{x}},\sqrt{J_{y}}\right)$
and (b) plot of lost particles versus $\sqrt{J_{x}+J_{y}}$ for different
longitudinal action.\label{fig:h-dist} The axis variables are normalized
by rms size of transverse action.}
\end{figure}

Beam emittance growth and loss rate are routinely measured during
the Tevatron operation. They can be directly compared with numerical
simulations but only for relatively short times. Figure \ref{fig:tev-loss}
(a) shows the time evolution of the four-dimensional emittance of
bunches \#1, \#6, and \#12 for 15 hours of high energy physics (HEP)
run of store \# 7650. %
 The emittance is calculated and plotted by $\epsilon_{4d}=\sqrt{\epsilon_{x}\epsilon_{y}}$.
It is observed that during the HEP run, the emittance growth is nearly
linear. The growth rate is 6.7\%/hr. Figure \ref{fig:tev-loss} (b)
shows the measured beam loss rates of anti-proton bunches during the
first 1 hour of store \#7601-\#7650 at collision energy 960 GeV. In
order to see the effects of beam-beam interactions on the beam loss,
the loss rate is obtained by subtracting the particle losses due to
luminosity at the main interaction points from the total beam loss
rate. Averaged loss rates of bunch \#1 and \#12 are 1.4 \%/hr and
1.2 \%/hr respectively, while the loss rate of bunch \#6 is 2.3 \%/hr.
We performed the simulations of emittance growth and particle loss
of anti-proton beam, as shown in Fig. \ref{fig:tev-loss} (c)-(d).
The particle tracking is carried out over $10^{7}$ turns corresponding
to approximately 3.5 minutes storage time of the Tevatron. In the
simulation, nominal tune is (20.571, 20.569). Initial transverse (95\%
normalized) emittance of anti-protons $\left(\epsilon_{x},\epsilon_{y}\right)$
is set to be (9.0,7.8) mm-mrad from averaging the measured emittances
while proton's initial emittance is (18,23) mm-mrad. Bunch intensities
of anti-proton and proton are $0.86\times10^{11}$ and $2.64\times10^{11}$
respectively. Figure \ref{fig:tev-loss} (c) shows the emittance growth
of three bunches during the simulation. The growth rate is approximately
9 \%/hr, which is close to the measured growth rate 7 \%/hr in Fig.
\ref{fig:tev-loss} (a). The emittance does not vary from bunch to
bunch. However, the beam losses vary considerably from bunch to bunch.
As shown in Fig. \ref{fig:tev-loss} (d), bunch \#6 loses more particles
than bunches \#1 and \#12, which agrees well with the observation.
For the simulation of beam loss, we used the hollow Gaussian distribution
in transverse action coordinates. Most of the lost particles have
large transverse actions as shown in Fig. \ref{fig:h-dist} (a), while
the lost particles are distributed over the entire range of longitudinal
action, as shown in Fig. \ref{fig:h-dist} (b). %

The compensation of long-range effects in the Tevatron with a current-carrying
wire was investigated using an earlier version of the code \citep{tsen-2}.
It was found that a single wire was unable to compensate for all the
70 interactions, since they were all at different betatron phases
from the wire.

\subsection{Relativistic Heavy Ion Collider}

\begin{figure}
\begin{centering}
\subfloat[]{\begin{centering}
\includegraphics[bb=20bp 5bp 340bp 210bp,clip,scale=0.7]{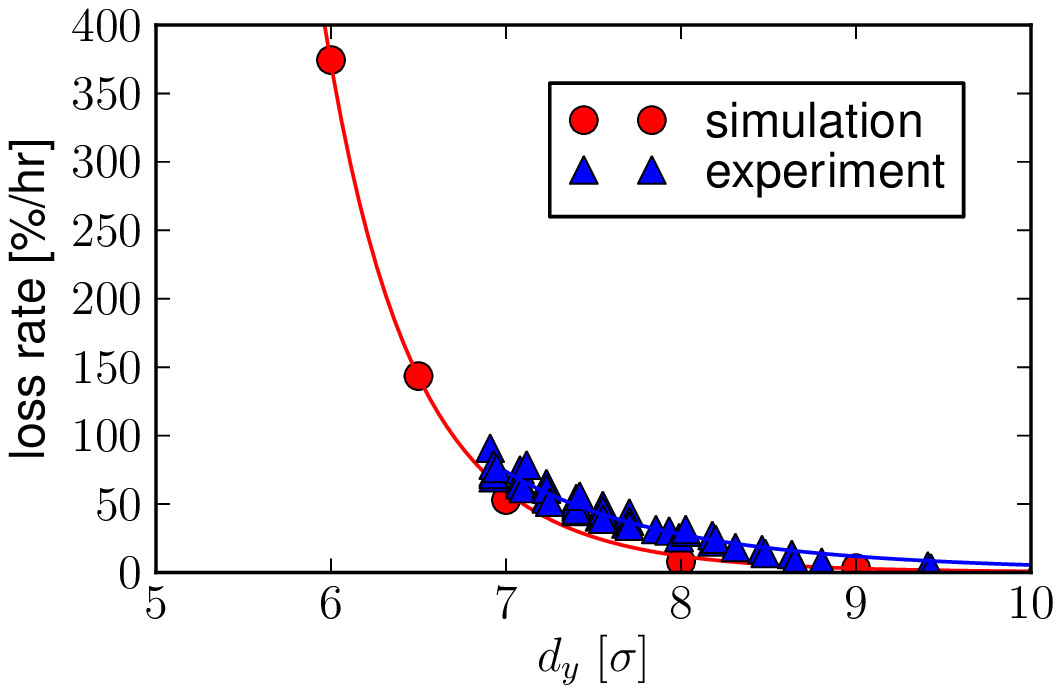}
\par\end{centering}

}\subfloat[]{\begin{centering}
\includegraphics[bb=20bp 5bp 340bp 210bp,clip,scale=0.7]{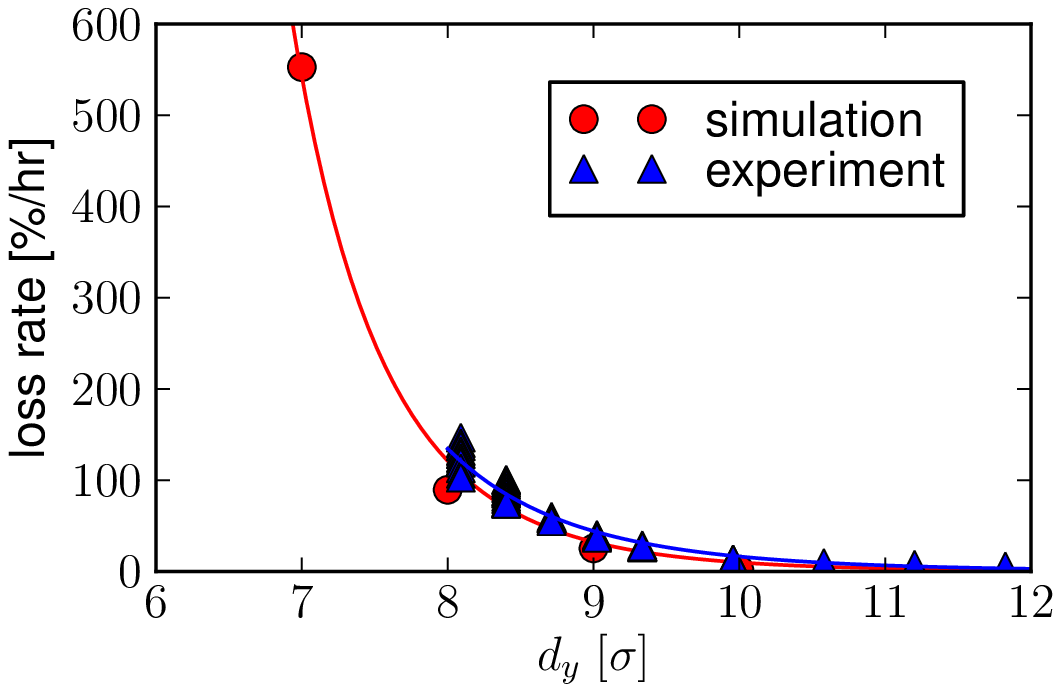}
\par\end{centering}

}
\par\end{centering}

\caption{Comparison of the simulated beam loss rates with the measured as a
function of separations. (a) gold beam at collision energy, (b) deuteron
beam at collision energy \citep{hjkim-1}. \label{fig:rhic-loss}}
\end{figure}

We have studied the effects of a current-carrying wire on the beam
dynamics in RHIC \citep{hjkim-1}. Two current-carrying wires, one
for each beam, have been installed between the magnets $Q3$ and $Q4$
of IP6 in the RHIC tunnel. In the physics run 9, an attempt was made
to compensate the long range beam-beam interaction which shows the
reduction of beam loss \citep{Rama}. During the physics run 7 and
8, the impact of current-carrying wires on a beam was measured without
an attempt to compensate the beam-beam interactions. However, the
experimental results help to understand the beam-beam effects because
the wire force is similar to the long-range beam-beam force at large
separations. As an example, Fig. \ref{fig:rhic-loss} plots the beam
loss rate due to the wire as a function of beam-wire separation distance.
 The onset of beam losses is observed at 8 $\sigma$ and 9 $\sigma$
for gold and deuteron beams, respectively. The threshold separation
for the onset of sharp losses observed in the measurements and simulations
agree to better than 1 $\sigma$. It is also significant that the
simulated loss rates at 7 and 8 $\sigma$ separation for the gold
beam and 8 and 9 $\sigma$ for the deuteron beam are very close to
the measured loss rates. At fixed separation, the wire causes a much
higher beam loss with the deuteron beam than with the gold beam. The
loss rate for the gold beam at a 8 $\sigma$ separation is about 10
\%/hr while for the deuteron beam the loss rate is about an order
of magnitude higher both in measurements and simulation. Simulations
of the beam loss rate when the wire is present are in good agreement
with the experimental observations.

In the proton-proton runs of RHIC, the maximum beam-beam parameter
reached so far is about $\xi=0.008$. This tune shift is large enough
that the combination of beam-beam and machine nonlinearities excite
betatron resonances which cause emittance growth and diffuse particles
into the tail of beam distribution and beyond. Consequently RHIC is
actively developing an electron lens for compensating the head-on
interactions \citep{Fischer-1}. In order to seek the electron lens
parameters at which the beam life time is improved, we choose three
different electron beam distribution functions: (a) $1\sigma$ Gaussian
distribution with the same rms beam size as that of the proton beam
$\sigma$, (b) $2\sigma$ Gaussian distribution with rms size twice
that of the proton beam, and (c) Smooth-edge-flat-top (SEFT) distribution
with an edge around at 4 $\sigma$. When the electron beam profile
matches the proton beam, the full compression of the tune spread requires
the electron beam intensity $N_{e}=4\times10^{11}$ which is defined
as the electron beam intensity required for full compensation. Table
\ref{tab:summary} shows the results of particle loss for different
intensities with the three electron beam profiles. %
\begin{table}
\begin{centering}
\begin{tabular}{c|c|c}
\hline 
Profile  & Intensity $\left(4\times10^{11}\right)$  & Particle loss$^{\dagger}$(\%)\tabularnewline
\hline 
$1\sigma$ Gaussian  & 1  & 635\tabularnewline
 & 1/2  & 115\tabularnewline
 & 1/4  & 63\tabularnewline
 & 1/8  & 30\tabularnewline
\hline 
$2\sigma$ Gaussian  & 4  & 93\tabularnewline
 & 2  & 10\tabularnewline
 & 1  & 8\tabularnewline
 & 1/2  & 6\tabularnewline
\hline 
SEFT  & 8  & 330\tabularnewline
 & 4  & 21\tabularnewline
 & 2  & 22\tabularnewline
 & 1  & 6\tabularnewline
 & 1/2  & 6\tabularnewline
\hline 
\multicolumn{3}{l}{$^{\dagger}$relative to that without beam-beam compensation}\tabularnewline
\end{tabular}
\par\end{centering}

\caption{Comparison of particle loss for different electron beam profiles and
intensities. \label{tab:summary}}
\end{table}

At an intensity $N_{e}=4\times10^{11}$, the particle loss is nearly
six times the loss without beam-beam compensation. The beam lifetime
at $N_{e}=2\times10^{11}$ however is comparable with that of no beam-beam
compensation. As the electron beam intensity is decreased, the particle
loss decreases significantly, and is reduced to 30\% of that without
beam-beam compensation at $N_{e}=0.5\times10^{11}$. For the $2\sigma$
Gaussian and SEFT electron beam profiles, we calculated particle loss
for different electron beam intensities. The upper limits of the electron
beam intensity for these two distributions are chosen so that peak
of the electron profile matches that of the full compensation at $1\sigma$
Gaussian. For the intensities $2\times10^{11}$ and $4\times10^{11}$
of $2\sigma$ Gaussian profile, there is a significant reduction in
beam loss, for example, below 10\% of the particle loss without beam-beam
compensation when the electron beam intensity is $2\times10^{11}$.
A significant improvement of beam lifetime with the SEFT profile is
also observed below $8\times10^{11}$. There is a threshold electron
beam intensity below which beam life time is increased: $2\times10^{11}$
for the $1\sigma$ Gaussian, $8\times10^{11}$ for the $2\sigma$
Gaussian, and $16\times10^{11}$for the SEFT profile. Particle loss
is relatively insensitive to electron lens current variations below
the threshold current with the $2\sigma$ Gaussian and SEFT profiles.
This looser tolerance on the allowed variations in electron intensity
will allow greater intensity fluctuations and is likely to be beneficial
during experiments.

\subsection{Large Hadron Collider}

\begin{figure}
\begin{centering}
\subfloat[]{\begin{centering}
\includegraphics[bb=20bp 5bp 340bp 210bp,clip,scale=0.7]{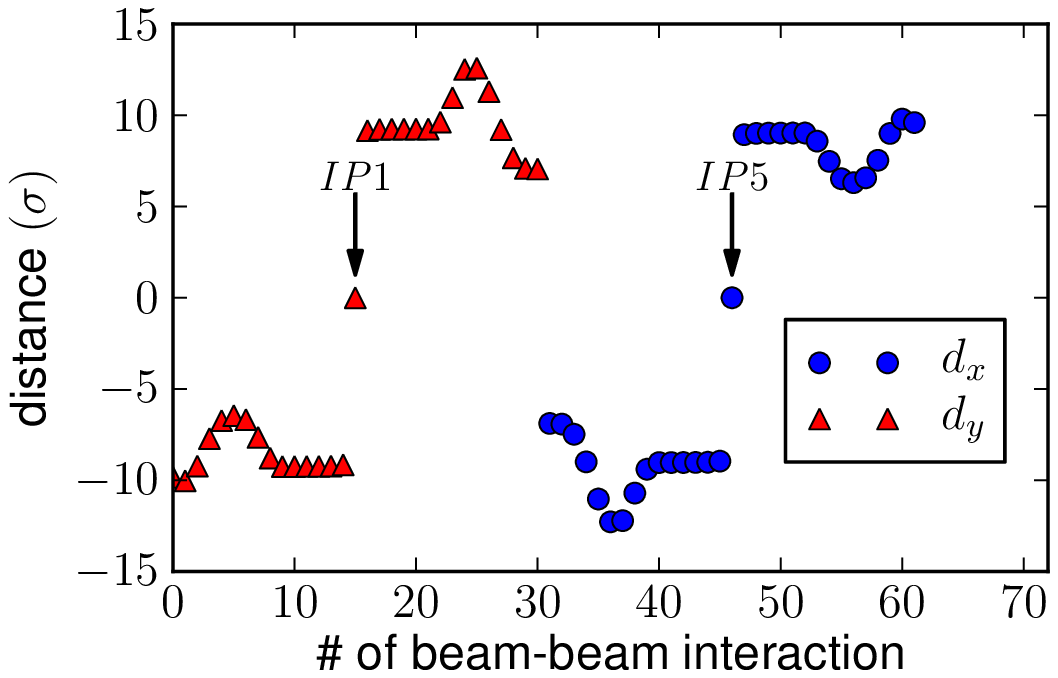}
\par\end{centering}

}\subfloat[]{\begin{centering}
\includegraphics[bb=20bp 5bp 340bp 210bp,clip,scale=0.7]{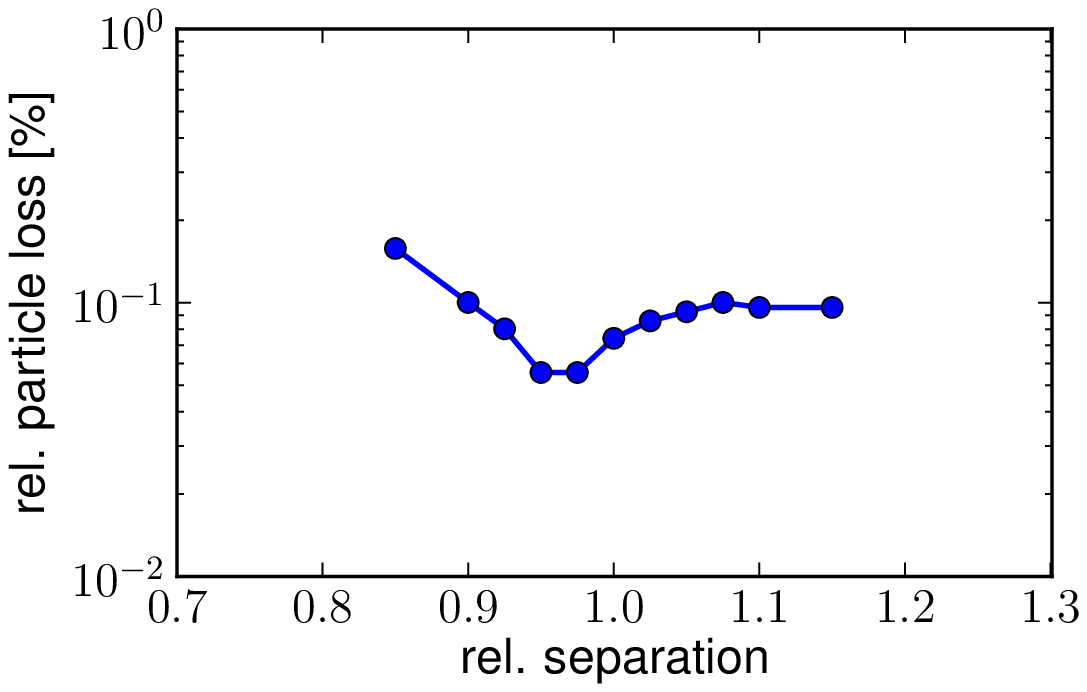}
\par\end{centering}

} 
\par\end{centering}

\caption{Plot of (a) beam-beam separation at IP 1 and 5 and (b) particle loss
according to wire separation distance with wire strength 82.8 Am.
\label{fig:lhc}}
\end{figure}

As mentioned above, long-range beam-beam interactions cause emittance
growth or beam loss in the Tevatron and are expected to deteriorate
beam quality in the LHC. Increasing the crossing angle to reduce their
effects has several undesirable effects, the most important of which
is a lower luminosity due to the smaller geometric overlap. For the
LHC, a wire compensation scheme has been proposed to compensate the
long-range interactions \citep{Koutchouk}. However, several issues
need to be resolved for efficient compensation. With the design bunch
spacing, there are about 30 long-range interactions on both sides
of an interaction point (IP). The beam-beam separation distance varies
from 6.3 $\sigma$ to 12.6 $\sigma$. The resulting beam-beam force
is not identical to that generated by a single or multiple wire(s)
but can be closely approximated by the wires. Unlike the Tevatron,
the long-range forces in the LHC are all at nearly the same betatron
phase and this makes the compensation scheme feasible. The wire-beam
separation distance is one of the parameters which determine the performance
of a wire compensator. Figure \ref{fig:lhc} (a) shows the beam-beam
separation distance normalized by the transverse rms bunch size. Two
counter-rotating beams collide at a vertical crossing angle near IP1
while they collide at a horizontal crossing angle near IP5. The separations
are asymmetric with respect to the interaction points. The reference
wire-beam separation (9 $\sigma$) is chosen as the average of beam-beam
separations. %
 Figure \ref{fig:lhc} (b) shows the results of particle loss for
different wire-beam separations. The particle loss saturates at large
separation while there is a sharp increase of particle loss at small
separation. We directly see the minimum particle loss between 0.9
and 1.0 of the reference separation. It reveals that the average of
beam-beam separations is close to an optimal separation between the
wire and the high energy bunch.

\section{Summary \label{sec:Summary}}

In this paper, an efficient parallel beam simulation model for circular
colliders is presented in order to study the effects of beam-beam
interactions and machine nonlinearities, and the effectiveness of
beam-beam compensation schemes. We have included the major nonlinearities
present in accelerators in our program as well as models for several
methods to compensate the effects of beam-beam interactions. A particle-domain
decomposition scheme is implemented with the master/slave configuration
to achieve a balanced workload in a parallel environment. A performance
test of beam-beam interactions indicates that the parallelization
scheme scales linearly in both the number of processors and the number
of particles in the beam. We have used the program to study the emittance
growth and beam loss of different bunches due to the beam-beam interactions
in the Tevatron, the compensation of head-on beam-beam interactions
with a low energy electron beam in RHIC, and the long-range beam-beam
compensation using a current-carrying wire in the Tevatron, RHIC and
the LHC. The pattern of beam losses observed in the Tevatron is reproduced
in the simulations. In RHIC, simulations of the beam loss rate when
the wire is present are in good agreement with the experimental observations.
We have several predictions from the results of head-on compensation
in RHIC. For example we find that proton beam life time is increased
if the electron beam intensity is kept below a threshold intensity.
An electron beam wider than the proton beam at the electron lens location
is found to increase beam life time. The results of LHC simulation
with the current carrying wire show that the particle loss is minimized
when the beam-wire separation is close to the average of beam-beam
separations.

\section{Acknowledgments}

We thank V. Boocha, B. Erdelyi and V. Ranjbar for their contributions
to the development of \texttt{BBSIM}. This research used resources
of the Accelerator Physics Center at Fermi National Accelerator Laboratory
as well as resources of the National Energy Research Scientific Computing
Center at Lawrence Berkeley National Laboratory, which is supported
by the Office of Science of the U.S. Department of Energy. This work
is partially supported by the US Department of Energy through the
US LHC Accelerator Research Program (LARP). Fermi National Accelerator
Laboratory (Fermilab) is operated by Fermi Research Alliance, LLC
under Contract No. DE-AC02-07CH11359 with the United States Department
of Energy.

\appendix
\bibliographystyle{elsarticle-num}
\bibliography{nima-bbsim_hjkim}

\end{document}